\documentclass[pra, aps, english, twocolumn, hyperref,floatfix, superscriptaddress,showpacs]{revtex4}
\usepackage{graphicx}
\usepackage{amssymb, amsmath, amsfonts, color, rotating, multirow, graphicx, bm}
\usepackage[dvipdfm,bookmarks=false,pdfstartview=FitH,hyperindex=true, colorlinks, linkcolor=blue, citecolor=blue]{hyperref}

\begin{document}
\author{Ran Wei}
\affiliation{Hefei National Laboratory for Physical Sciences at Microscale and Department
of Modern Physics, University of Science and Technology of China, Hefei, Anhui 230026, China}
\affiliation{Laboratory of Atomic and Solid State Physics, Cornell University, Ithaca, NY, 14850}
\author{Erich J. Mueller}
\affiliation{Laboratory of Atomic and Solid State Physics, Cornell University, Ithaca, NY, 14850}
\title{Magnetic field dependence of Raman coupling in Alkali atoms}
\date{\today}
\begin{abstract}
We calculate the magnetic field dependence of Rabi rates for two-photon optical Raman processes in alkali atoms.
Due to a decoupling of the nuclear and electronic spins, these rates fall with increasing field.
At the typical magnetic fields of alkali atom Feshbach resonances ($B\sim 200$G$-1200$G),
the Raman rates have the same order of magnitude as their zero field values, suggesting one can combine Raman-induced
gauge fields or spin-orbital coupling with strong Feshbach-induced interactions.
The exception is $^6$Li, where there is a factor of $7$ suppression in the Raman coupling,
compared to its already small zero-field value.
\end{abstract}
\maketitle

\section{introduction}
Two-photon ``Raman" transitions act as an important control parameter in cold atom
experiments. These optical transitions couple motional and internal degrees of freedom, mimicking important physical processes
such as gauge fields \cite{gaugeboson,lattice} and spin-orbital couplings \cite{SOboson,PanSO,SOshanxi,SOmit}. They have also been
used as spectroscopic probes, for example allowing scientists to measure excitation spectra
\cite{Dalibard2007,Zhang2012,Zhang2012p}. The most exciting future applications
of these techniques will involve strongly interacting atoms near Feshbach resonances
\cite{Sarma2008,Fujimoto2009,Duan2011,Zoller2011,Cooper2011,Chuanwei2012,Huhui2012,Huhui2012r,Liu2012,Mueller2012}.
Here we study the Raman couplings as a function of magnetic field, quantifying the practicality of
such experiments. We find that for relatively heavy atoms, such as $^{40}$K, the Raman techniques are
compatible with the magnetic fields needed for Feshbach resonances. Despite important experimental demonstrations \cite{SOmit},
lighter atoms, such as $^6$Li are less promising, as the ratio of Raman Rabi frequency to the inelastic scattering rate
is not sufficiently large. This problem is exacerbated by the magnetic field \cite{suppress}.

We are interested in two-photon transitions which take an atom between two hyperfine states $|g_1\rangle$
and $|g_2\rangle$. Optical photons only couple to electronic motion in an atom.
Hence such Raman transitions rely on fine and hyperfine interactions. The former couple electronic spins and motion, and
the latter couple nuclear spins to the electronic angular momentum.
One expects that magnetic fields will reduce these Raman matrix elements, as the disparate Zeeman coupling
of electronic and nuclear degrees of freedom competes with the fine and hyperfine interactions.
We find that at the typical magnetic fields of Feshbach resonance ($B\sim 200$G$-1200$G \cite{Chengchin2010}),
the Raman couplings are still quite strong. The exception is $^6$Li, where there is a factor of $7$ suppression,
compared to its already small zero-field value.

A key figure of merit in experiment is the ratio of the Raman Rabi frequency to the inelastic scattering rate
$\beta\equiv\Omega_R/\Gamma_{\rm ine}$. The inverse Rabi frequency gives the time required for an atom to flip between
$|g_1\rangle$ and $|g_2\rangle$, and the inverse inelastic scattering rate gives the average time between photon absorption events,
which leads to heating. For equilibrium experiments on Raman-dressed atoms, one needs $\Omega_R\sim\mu/\hbar\sim{\rm kHz}$,
where $\mu$ is the chemical potential. A typical experiment takes one second.
Thus if $\beta\lesssim10^3$, the inelastic light scattering has a large impact.
As argued by Spielman \cite{Spielman2009}, both of these rates at sufficiently large detuning limit
are proportional to the laser intensity and inversely proportional to the square of the detuning.
We numerically calculate this ratio as a function of magnetic field,
including all relevant single-particle physics. We also explore
the detuning dependence of this ratio.

The remainder of this manuscript is organized as follows. In Sec. II,
we estimate the ratio of the Raman Rabi frequency to
the inelastic scattering rate in the absence of magnetic field for various alkali atoms.
In Sec. III, we calculate the magnetic field dependence of electric dipole transitions:
We introduce the single-particle Hamiltonian in Sec. III(A),
and in Sec. III(B) we introduce the formal expression of the electric dipole transitions.
The analytical discussions and numerical results are elaborated in Sec. III(C) and Sec. III(D).
In Sec. IV, we calculate the ratios for $^{23}$Na, $^{40}$K, $^{85}$Rb, $^{87}$Rb and $^{133}$Cs,
and further study $^6$Li and explore how the ratio depends on detuning.
Finally we conclude in Sec. V.

\section{Raman coupling}

\begin{figure}[!htb]
\includegraphics[width=5cm]{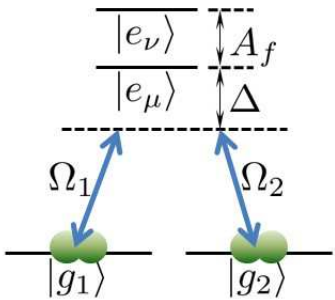}
\caption{(Color online) Sketch of the energy level structures in a Raman experiment.
The Rabi frequencies $\Omega_1,\Omega_2$ characterize the coupling strengths
between the ground states $|g_1\rangle,|g_2\rangle$ and the excited states.
The fine structure energy splitting between $|e_\mu\rangle$ and $|e_\nu\rangle$ is $A_f=E_\nu-E_\mu$.
The laser detuning is $\Delta=(E_\mu-E_g)-\hbar\omega$.
\label{raman}}
\end{figure}

We consider a typical setup of a Raman experiment, as shown in Fig. \ref{raman}:
Two hyperfine ground states $|g_1\rangle$ and $|g_2\rangle$ with energies $E_g$,
are coupled to a pair of excited multiplets $\{|e_\mu\rangle,|e_\nu\rangle\}$ by two lasers,
where the coupling strengths are characterized by the Rabi frequencies $\Omega_1$ and $\Omega_2$.
The states $|e_\mu\rangle$ and $|e_\nu\rangle$ are states in the
$J=1/2$ and $J=3/2$ manifolds, with energies $E_\mu$ and $E_\nu$,
and the energy difference $A_f=E_\nu-E_\mu$. For $^6$Li, $A_f\sim(2\pi\hbar)\times10$GHz.
For heavier atoms such as $^{40}$K, $A_f\sim(2\pi\hbar)\times1$THz.
The laser detuning $\Delta=(E_\mu-E_g)-\hbar\omega$, characterizes the energy mismatch between the laser
frequency $f=\omega/2\pi$ and the atomic ``D1" transition.

Within such a setup, the two lasers couple $|g_1\rangle$ and $|g_2\rangle$ via two-photon transitions,
and the Raman Rabi frequency, which characterizes the (Raman) coupling strength, is
\begin{eqnarray}
\label{ramaneq}
\Omega_R=\sum_{\mu}\frac{\hbar\Omega_{1\mu}\Omega_{2\mu}}{4\Delta}
+\sum_{\nu}\frac{\hbar\Omega_{1\nu}\Omega_{2\nu}}{4(\Delta+A_f)},
\end{eqnarray}
where the optical Rabi frequency $\Omega_{i\epsilon}=\frac{{\bm E}_i\cdot\langle g_i|{\bm d}|e_{\epsilon}\rangle}{\hbar}$
with electronic dipole ${\bm d}=e{\bm r}$, characterizes the individual transition element
between the ground state $|g_i\rangle$ and the excited state $|e_\epsilon\rangle$ ($i=1,2$ and $\epsilon=\mu,\nu$).
In our following calculations, we assume $|g_1\rangle$ and $|g_2\rangle$ are the lowest two ground states.

This expression can be simplified by noting that the ground state quadrupole matrix element
$\langle g_i|d_ad_b|g_j\rangle=0$ unless $i=j$ and $a=b$ ($a,b=x,y,z$).
This reflects the spherical symmetry of the electron wavefunction, and the fact that the electronic dipole
does not couple to spin. Inserting a complete set of exited states, we find
$\sum_\mu\Omega_{1\mu}\Omega_{2\mu}+\sum_\nu\Omega_{1\nu}\Omega_{2\nu}=0$,
allowing us to write Eq. (\ref{ramaneq}) solely in terms of the matrix elements for the D1 line,
\begin{eqnarray}
\Omega_R=\frac{\hbar A_f}{4\Delta(\Delta+A_f)}\sum_{\mu}\Omega_{1\mu}\Omega_{2\mu}.
\end{eqnarray}
It is thus clear that $\Omega_R\sim A_f/\Delta^2$ for $\Delta\gg A_f$.
The inelastic scattering rate that emerges from the spontaneous emission of the excited states is
\begin{eqnarray}
\label{inelasticeq}
\Gamma_{\rm{ine}}=\gamma\left(\sum_{\mu}\frac{\hbar^2(\Omega_{1\mu}^2+\Omega_{2\mu}^2)}{4\Delta^2}
+\sum_{\nu}\frac{\hbar^2(\Omega_{1\nu}^2+\Omega_{2\nu}^2)}{4(\Delta+A_f)^2}\right)
\end{eqnarray}
where $\gamma$ denotes the decay rate of these excited states.
For $\Delta\gg A_f$, this rate scales as $\Gamma_{\rm{ine}}\sim\gamma/\Delta^2$.
Explicit calculations show that to a good approximation
$\beta\equiv\Omega_R/\Gamma_{\rm ine}\approx\beta_e\equiv A_f/12\hbar\gamma$
for $\Delta\gg A_f$. The factor of $\frac{1}{12}$ can crudely be related to cancellation of
terms of opposite signs in the expression for $\Omega_R$.
As an illustration, we show $A_f$, $\gamma$, and $\beta_e$ for various alkali atoms in Tab. I.
We also present results of our numerical calculation of $\beta$. The details of this calculation will be given in Sec. III.
\begin{table}[ht]
\centering
\begin{tabular}{| c | c | c | c | c | c | c | c |}
\hline
Alkalis & $^6$Li($2p$) & $^6$Li($3p$) & $^{23}$Na & $^{40}$K & $^{85}$Rb & $^{87}$Rb & $^{133}$Cs\\
\hline
$A_f/$($2\pi\hbar$)GHz & $10.0$ & $2.88$ & $515$ & $1730$ & $7120$ & $7120$ & $16600$\\
\hline
$\gamma/$($2\pi$)MHz & $5.87$ & $0.754$ & $9.76$ & $6.04$ & $5.75$ & $5.75$ & $4.57$\\
\hline
$\beta_e/10^{3}$  & $0.14$ & $0.32$ & $4.4$ & $24$ & $103$ & $103$ & $303$ \\
\hline
$\beta/10^{3}$  & $0.13$ & $0.30$ & $4.0$ & $23$ & $101$ & $103$ & $304$ \\
\hline
\end{tabular}
\caption{Fine structure energy splitting $A_f$, spontaneous decay rate $\gamma$,
and ratios $\beta_e$ and $\beta$ for various alkali atoms.
For $^6$Li, we consider either $2p$ states or $3p$ states as the excited multiplet. For other atoms,
we consider the lowest $p$ multiplet. The ground states for all alkali atoms
are the two lowest magnetic substates.
Data in the first two rows were extracted from archived data \cite{archive}.}
\end{table}

As seen from the table, the heavier atoms have more favorable ratios
($\beta\gtrsim 10^3$ for most alkali atoms).
For $^6$Li we include the rates for Raman lasers detuned from the
$2s-2p$ line and the narrower $2s-3p$ line. For all other atoms we just consider
the lowest energy $s-p$ transition.
We see that the ratio for $^6$Li can be improved by a factor of $2.2$ by using the $3p$ states.
Similar gains are found for laser cooling schemes using these states \cite{Hulet2011}.

\section{magnetic field dependence of electric dipole transitions}
In this section we will calculate $\Omega_{i\epsilon}$ and its dependence on the magnetic field.
In the following section we will use these results to calculate $\Omega_R$ and $\Gamma_{\rm ine}$.

\subsection{Single-particle Hamiltonian}
Fixing the principle quantum number of the valence electron, the fine and hyperfine atomic structure
of an alkali in a magnetic field is described by a coupled spin Hamiltonian
\begin{eqnarray}
\label{ham1}
H&=&H_a+H_B
\end{eqnarray}
where
\begin{eqnarray}
\label{ham1a}
H_a&=&c_f{\bm L}\cdot{\bm S}+c_{hf1}{\bm L}\cdot{\bm I}+c_{hf2}{\bm S}\cdot{\bm I}\\
\label{ham1b}
H_B&=&\mu_B(g_L{\bm L}+g_S{\bm S}+g_I{\bm I})\cdot {\bm B}.
\end{eqnarray}
Here the vectors ${\bm L}$ and ${\bm S}$ are the dimensionless orbital and spin angular momentum of the electron,
and ${\bm I}$ is the angular momentum of the nuclear spin. The coefficients $c_f$ and $c_{hf1},c_{hf2}$
are the fine structure constant and hyperfine structure constants, which were measured in experiments \cite{hyperfine,archive}.
$\mu_B$ is the Bohr magneton and $g_L,g_S,g_I$ are the Lande $g$-factors.

For $^6$Li, the Zeeman splitting energy is $E_B\sim(2\pi\hbar)\times5$GHz at the
magnetic field of the wide Feshbach resonance $B=834$G \cite{Chengchin2010}.
This splitting is comparable to the fine structure constant $c_f\sim(2\pi\hbar)\times7$GHz.
For other heavier atoms where $E_B\ll c_f\sim(2\pi\hbar)\times1$THz,
the fine structure interaction is robust against the magnetic field, and we can appropriate the Hamiltonian as
\begin{eqnarray}
\label{ham2}
H=c_{hf}{\bm J}\cdot{\bm I}+\mu_B(g_J{\bm J}+g_I{\bm I})\cdot {\bm B}
\end{eqnarray}
with the vector ${\bm J}={\bm L}+{\bm S}$.

The Hamiltonian (\ref{ham1}) can be diagonalized in the
basis $|Lm_Lm_Sm_I\rangle$, where $\{m_L,m_S,m_I\}$ are the $z$-components of $\{L,S,I\}$,
and the eigenstate $|LQ\rangle$ can be expanded as
\begin{eqnarray}
\label{eigenstate}
|LQ\rangle=\sum_{m_Lm_Sm_I}C^Q_{m_Lm_Sm_I}|Lm_Lm_Sm_I\rangle
\end{eqnarray}
where $Q$ labels the eigenstate, and $C^Q_{m_Lm_Sm_I}$ corresponds to the eigenvector.
We will use these coefficients to calculate the electric dipole transition in the following
subsections.

\subsection{Formal expressions}
We define $D_q\equiv\langle Lm_Lm_Sm_I|er_q|L'm'_Lm'_Sm'_I\rangle$, the electric dipole transition
between $|Lm_Lm_Sm_I\rangle$ and $|L'm_L'm_S'm_I'\rangle$,
where $r_q$ is the position operator, expressed as an irreducible spherical tensor: $q=-1,0,1$ correspond
to $\sigma^-,\pi,\sigma^+$ polarized light.
Note that the electric dipole $er_q$ does not directly couple to the electric spin $m_S$ or the nuclear spin $m_I$, and
$D_q$ is of the form $D_q=\delta_{m_Sm'_S}\delta_{m_Im'_I}\langle Lm_L|er_q|L'm'_L\rangle$.
Using the Wigner-Eckart theorem, we obtain
\begin{eqnarray}
\label{dipole}
D_q=\delta_{m_Sm'_S}\delta_{m_Im'_I}W_{m_L'qm_L}^{L'L}
    \langle L||er||L'\rangle
\end{eqnarray}
where $\langle L||er||L'\rangle$ is the reduced matrix element, independent of $\{m_L,m_S,m_I\}$.
The coefficient $W_{m_L'qm_L}^{L'L}$ can be
written in terms of the Wigner $3$-$j$ symbol \cite{Brink}
\begin{eqnarray}
W_{m_L'qm_L}^{L'L}=(-1)^{L'-1+m_L}\sqrt{2L+1}\left(\begin{array}{ccc}
    L' & 1 & L\\
    m_L' & q & -m_L\\
    \end{array}\right)
\end{eqnarray}
Combining Eq. (\ref{eigenstate}) and Eq. (\ref{dipole}),
we obtain the electric dipole transition between two eigenstates $|LQ\rangle$ and $|L'Q'\rangle$,
\begin{eqnarray}
&&D_{q,LQ}^{L'Q'}\equiv\langle LQ|er_q|L'Q'\rangle\\
\notag&=&\sum_{\bar m_L\bar m'_L\bar m_S\bar m_I}C^{Q}_{\bar m_L\bar m_S\bar m_I}
C^{Q'}_{\bar m'_L\bar m_S\bar m_I}W_{\bar m_L'q\bar m_L}^{L'L}\langle L||er||L'\rangle
\end{eqnarray}
where the coefficients $C^{Q}_{\bar m_L\bar m_S\bar m_I}$ are defined by Eq. (\ref{eigenstate}).

\subsection{Analytical discussions}
While $C^{Q}_{\bar m_L\bar m_S\bar m_I}$ can be numerically calculated by extracting the
eigenvector of the Hamiltonian, in some regimes the problem simplifies, and
$C^{Q}_{\bar m_L\bar m_S\bar m_I}$ corresponds to a Clebsch-Gordan coefficient.
In this subsection we discuss these simple limits.

In a weak magnetic field such that the Zeeman splitting energy $E_B\ll c_{hf}$,
the electronic angular momentum and the nuclear spins are strongly mixed.
Here $Q$ corresponds to the three quantum numbers $\{J,F,m_F\}$,
where $F$ is the quantum number associated with the total hyperfine spin
${\bm F}={\bm J}+{\bm I}$, and $m_F$ labels the magnetic sublevels.
In this limit $C^{Q}_{\bar m_L\bar m_S\bar m_I}=C^{JFm_F}_{\bar m_L\bar m_S\bar m_I}
=\langle \bar m_L\bar m_S|J\bar m_J\rangle \langle \bar m_I\bar m_J|Fm_F\rangle$ is simply
the product of two Clebsch-Gordan coefficients.

In the regime $c_{hf}\ll E_B\ll c_f$,
the electronic angular momentum and the nuclear spins are decoupled, and $Q$ corresponds to $\{J,m_J,m_I\}$,
with $C^{Q}_{\bar m_L\bar m_S\bar m_I}=\langle \bar m_L\bar m_S\bar m_I|Jm_Jm_I\rangle
=\delta_{\bar m_Im_I}\langle \bar m_L\bar m_S|Jm_J\rangle$.
In this case, the dipole transition element obeys $D_{q,LQ}^{L'Q'}\propto\delta_{m_Im_I'}$,
and the states with different nuclear spins are not coupled by the lasers.

In an extremely strong magnetic field such that $E_B\gg c_f$,
the electronic spins and the electronic orbital angular momentum are decoupled, and $Q$ corresponds to $\{m_L,m_S,m_I\}$,
with $C^{Q}_{\bar m_L\bar m_S\bar m_I}=\delta_{\bar m_Lm_L}\delta_{\bar m_Sm_S}\delta_{\bar m_Im_I}$.
In this case, the dipole transition element obeys $D_{q,LQ}^{L'Q'}\propto\delta_{m_Sm_S'}\delta_{m_Im_I'}$,
and states with disparate nuclear or electronic spin projections are not coupled by the lasers.
In short, the large fields polarize the electronic spins and nuclear spins,
making them robust quantum numbers which cannot be influenced by optical fields.

\subsection{Numerical results}
Here we numerically calculate $D_{q,LQ}^{L'Q'}$ in the intermediate regime $E_B\gtrsim c_{hf}$.
For $^6$Li, $c_{hf}\lesssim E_B\lesssim c_f$, one needs to diagonalize the Hamiltonian (\ref{ham1}).
The numerics are simpler for other alkali atoms where $c_{hf}\lesssim E_B\ll c_f$.
In this case $Q$ is decomposed into $J$ and $\tilde Q$, with the latter labeling the eigenstates of the
simplified Hamiltonian in Eq. (\ref{ham2}).
The coefficient $C_{\bar m_L\bar m_S\bar m_I}^Q$ is then reduced as
$C_{\bar m_L\bar m_S\bar m_I}^Q=\langle \bar m_L\bar m_S|J\bar m_J\rangle C^{\tilde Q}_{\bar m_J\bar m_I}$.

\begin{figure}[!htb]
\includegraphics[width=7cm]{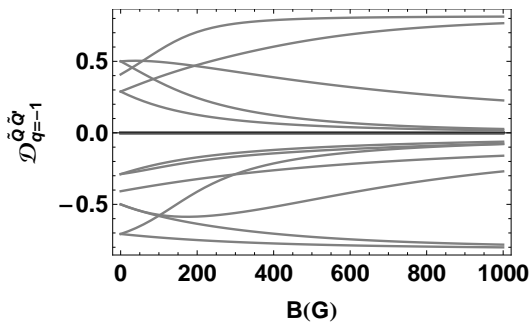}
\caption{(Color online) Dimensionless
electric dipole transition $\mathcal{D}_{q}^{\tilde Q\tilde Q'}\equiv
D_{q,LJ\tilde Q}^{L'J'\tilde Q'}/\langle L||er||L'\rangle$
as a function of magnetic field for $^{23}$Na, with the parameters $L=0,J=1/2,L'=1,J'=1/2$ and $q=-1$.
The twelve different lines correspond to all of the various allowed
dipole transitions for $\sigma^-$ light.
\label{coupling}}
\end{figure}

As an illustration, in Fig. \ref{coupling} we plot the dimensionless electric dipole transition
$\mathcal{D}_{q}^{\tilde Q\tilde Q'}\equiv
D_{q,LJ\tilde Q}^{L'J'\tilde Q'}/\langle L||er||L'\rangle$
as a function of the magnetic field for $^{23}$Na,
where we choose the eigenstates of $L=0,J=1/2$ and $L'=1,J'=1/2$ as the initial states
and the final states (D1 transitions), and use $\sigma^-$ polarized light.

The $F=1,2$ and $F'=1,2$ manifolds allow twelve $\sigma^-$ transitions.
At large fields, the absolute values of four of them saturate at finite values
while the rest of them approach zero.
This large field result stems from the decoupling of the electronic and nuclear spins.
(the $\tilde Q$ eigenstates can be described by $m_J,m_I$ in that limit).
Under such circumstance, the allowed transitions for the $\sigma^-$ transition
only occur at $m_J=1/2,m_J'=-1/2$ and $m_I=m_I'$ with $m_I=-3/2,-1/2,1/2,3/2$.

\section{Magnetic field dependance of Raman coupling}

We have illustrated how the electric dipole transition $\mathcal{D}_{q}^{\tilde Q\tilde Q'}$ depends on
the magnetic field. Here we calculate the ratio $\beta=\Omega_R/\Gamma_{\rm ine}$
using Eq. (\ref{ramaneq}) and Eq. (\ref{inelasticeq}) from Sec. II,
and the relation $\Omega_{i\epsilon}\propto\sqrt{I_i}\mathcal{D}_{q}^{\tilde Q\tilde Q'}$, with
$I_i$ the intensity of each laser.

\begin{figure}[!htb]
\includegraphics[width=7cm]{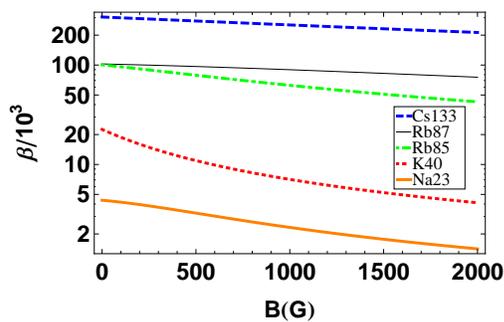}
\caption{(Color online) Ratio $\beta=\Omega_R/\Gamma_{\rm ine}$ as a function of magnetic field $B$ for various alkali
atoms at $\Delta=(2\pi\hbar)\times100$THz. $\Omega_R$ is the Raman Rabi rate, and $\Gamma_{\rm ine}$ is the
inelastic scattering rate. Note the logarithmic vertical scale.
\label{ratio}}
\end{figure}

In Fig. \ref{ratio}, we plot $\beta$ as a function of the magnetic
field for various alkali atoms at $\Delta=(2\pi\hbar)\times100$THz $\gg A_f$,
where we assume $|g_1\rangle$ and $|g_2\rangle$ are the lowest ground states,
and the two lasers with $\sigma^+$ and $\pi$ polarized light have an equal intensity.
We see that although the ratios decrease with the magnetic field,
they are still quite appreciable at $B\sim 200$G$-1200$G, suggesting that the Raman experiment and
strong Feshbach-induced interactions are compatible.

\begin{figure}[!htb]
\includegraphics[width=7cm]{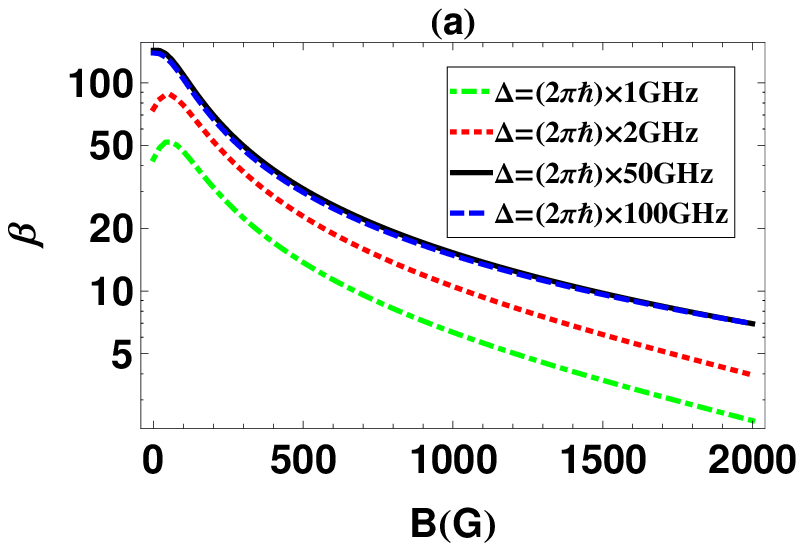}
\includegraphics[width=7cm]{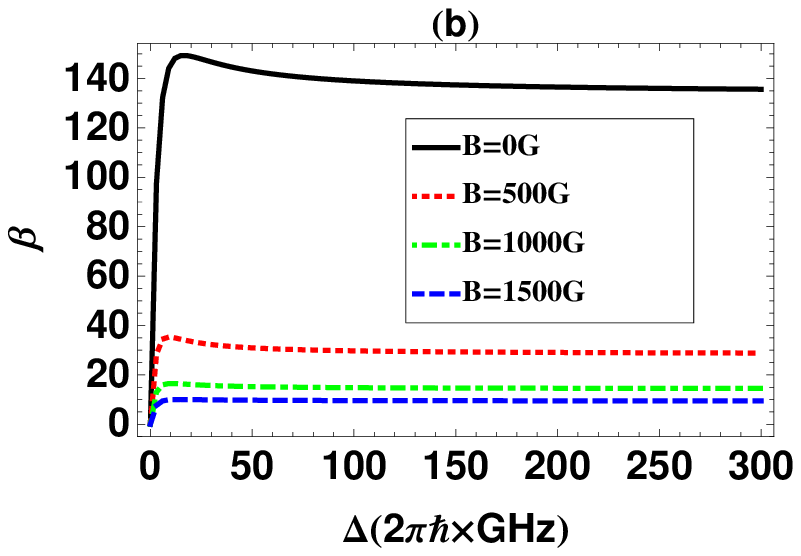}
\caption{(Color online) (a): Ratio $\beta=\Omega_R/\Gamma_{\rm ine}$
as a function of magnetic field $B$ for various detuning $\Delta$ for $^6$Li.
(b): Ratio $\beta$ as a function of detuning $\Delta$ for various magnetic field $B$.
\label{ratioLi6}}
\end{figure}

Given that $^6$Li is in a different regime than the other alkalis, it is
convenient to discuss its properties separately. Analyzing Eqs. (\ref{ham1a})-(\ref{ham1b}),
without making the approximations inherent in Eq. (\ref{ham2}), yields the transition rates in Fig. \ref{ratioLi6}.
Looking at the blue (dashed) curve in Fig. \ref{ratioLi6}(a), we see at the large detuning,
the ratio of the Raman Rabi rate to the inelastic scattering rate
for $^6$Li decreases faster than those for the heavier atoms in Fig. $\ref{ratio}$.
This rapid fall-off can be attributed to the much weaker coupling between the electronic
and nuclear spins in $^6$Li. At the magnetic field of Feshbach
resonance $B=834$G \cite{Chengchin2010}, $\beta$ is suppressed by a factor of $7$.
At small detuning $\Delta$ the ratio $\beta$ changes non-monotonically with the magnetic field,
as $\Gamma_{\rm{ine}}$ decreases faster than $\Omega_R$.
In Fig. \ref{ratioLi6}(b), we see that $\beta$ increases rapidly for small $\Delta$ and levels out at large $\Delta$.
There is an optimal detuning near $\Delta\approx A_f$, where $\beta$ has a maxima.
The peak value of $\beta$, however, is only marginally larger than its large $\Delta$ asymptotic value.
Moreover, the peak is further reduced as $B$ increases.

\section{conclusions}
In summary, we comprehensively studied the Raman
Rabi rates of alkali atoms in the presence of a magnetic field.
While the ratio of the Raman Rabi frequency to the inelastic scattering rate
decreases with the magnetic field, the suppression is {\emph {not}} significant
for most alkali atoms at the typical fields of the Feshbach resonance.
Our primary motivation is evaluating the feasibility of using Raman techniques
to generate strongly interacting Fermi gases with spin-orbit coupling. We conclude
that $^6$Li is not a good candidate, but $^{40}$K is promising.

\section{acknowledgement}
We thank Randall Hulet for many enlightening discussions.
R. W. is supported by CSC, CAS, NNSFC (Grant No.11275185),
and the National Fundamental Research Program (Grant No.2011CB921304).
This work was supported by the Army Research Office with funds from the DARPA
optical lattice emulator program.


\begin{thebibliography}{99}


\bibitem{gaugeboson}
Y.-J. Lin, R. L. Compton, K. Jim\'{e}nez-Garc\'{\i}a, J. V. Porto, and I. B. Spielman, Nature (London) \textbf{462}, 628 (2009).

\bibitem{lattice}
M. Aidelsburger, M. Atala, S. Nascimb\`{e}ne, S. Trotzky, Y.-A. Chen, and I. Bloch,
Phys. Rev. Lett. \textbf{107}, 255301 (2011).

\bibitem{SOboson}
Y.-J. Lin, K. Jim\'{e}nez-Garc\'{\i}a and I. B. Spielman, Nature (London) \textbf{471}, 83 (2011).

\bibitem{PanSO}
J.-Y. Zhang, S.-C. Ji, Z. Chen, L. Zhang, Z.-D. Du, B. Yan, G.-S. Pan, B. Zhao, Y.-J. Deng, H. Zhai, S. Chen, and J.-W. Pan,
Phys. Rev. Lett. \textbf{109}, 115301 (2012).

\bibitem{SOshanxi}
P. Wang, Z.-Q. Yu, Z. Fu, J. Miao, L. Huang, S. Chai, H. Zhai, and J. Zhang, Phys. Rev. Lett. \textbf{109}, 095301 (2012).

\bibitem{SOmit}
L. W. Cheuk, A. T. Sommer, Z. Hadzibabic, T. Yefsah, W. S. Bakr, and M. W. Zwierlein, Phys. Rev. Lett. \textbf{109}, 095302 (2012).

\bibitem{Dalibard2007}
T.-L. Dao, A. Georges, J. Dalibard, C. Salomon, and I. Carusotto, Phys. Rev. Lett. \textbf{98}, 240402 (2007).

\bibitem{Zhang2012}
P. Wang, Z. Fu, L. Huang, and J. Zhang, Phys. Rev. A \textbf{85}, 053626 (2012).

\bibitem{Zhang2012p}
Z. Fu, P. Wang, L. Huang, Z. Meng, and J. Zhang, Phys. Rev. A \textbf{86}, 033607 (2012).

\bibitem{Sarma2008}
C. Zhang, S. Tewari, R. M. Lutchyn, and S. DasSarma, Phys. Rev. Lett. \textbf{101}, 160401 (2008).

\bibitem{Fujimoto2009}
M. Sato, Y. Takahashi, and S. Fujimoto, Phys. Rev. Lett. \textbf{103}, 020401 (2009).

\bibitem{Duan2011}
S.-L. Zhu, L.-B. Shao, Z. D. Wang, and L.-M. Duan, Phys. Rev. Lett. \textbf{106}, 100404 (2011).

\bibitem{Zoller2011}
L. Jiang, T. Kitagawa, J. Alicea, A. R. Akhmerov, D. Pekker, G. Refael, J. I. Cirac,
E. Demler, M. D. Lukin, and P. Zoller, Phys. Rev. Lett. \textbf{106}, 220402 (2011).

\bibitem{Cooper2011}
N. R. Cooper, Phys. Rev. Lett. \textbf{106}, 175301 (2011).

\bibitem{Chuanwei2012}
M. Gong, G. Chen, S. Jia, and C. Zhang, Phys. Rev. Lett. \textbf{109}, 105302 (2012).

\bibitem{Huhui2012}
X.-J. Liu and H. Hu, Phys. Rev. A \textbf{85}, 033622 (2012).

\bibitem{Huhui2012r}
X.-J. Liu, L. Jiang, H. Pu, and H. Hu, Phys. Rev. A \textbf{85}, 021603(R) (2012).

\bibitem{Liu2012}
X.-J. Liu and P. D. Drummond, Phys. Rev. A \textbf{86}, 035602 (2012).

\bibitem{Mueller2012}
R. Wei and E. J. Mueller, Phys. Rev. A \textbf{86}, 063604 (2012).

\bibitem{suppress}
The subject of how magnetic fields suppress two-photon transitions has a long history,
some aspects of which are nicely recounted in Kassler's Nobel lectures \cite{Nobel}, where
he referred to the suppression of matrix element as a ``generalized Franck-Condon principle".

\bibitem{Nobel}
A. Kastler, Nobel lecture 195-196 (1966).

\bibitem{Chengchin2010}
C. Chin, R. Grimm, P. Julienne, and E. Tiesinga, Rev. Mod. Phys. \textbf{82}, 1225 (2010).

\bibitem{Spielman2009}
I. B. Spielman, Phys. Rev. A \textbf{79}, 063613 (2009).

\bibitem{archive}
The data for $^6$Li($2p$) was extracted from M. E. Gehm's PhD thesis:
Preparation of an optically trapped degenerate Fermi gas of $^6$Li:
finding the route to degeneracy (Duke, 2003).
The data of $^{40}$K was extracted from T. Tiecke's PhD thesis:
Feshbach resonances in ultracold mixtures of the fermionic gases $^6$Li and $^{40}$K (Amsterdam, 2010).
The data of $^{23}$Na, $^{85}$Rb, $^{87}$Rb, $^{133}$Cs was extracted
from D. A. Steck's online public resources (http://steck.us/alkalidata).
The data of $^6$Li($3p$) was from the private communications with R. G. Hulet (2012).

\bibitem{Hulet2011}
P. M. Duarte, R. A. Hart, J. M. Hitchcock, T. A. Corcovilos, T.-L. Yang, A. Reed, and R. G. Hulet,
Phys. Rev. A \textbf{84}, 061406(R) (2011).

\bibitem{hyperfine}
E. Arimondo, M. Inguscio, and P. Violino, Rev. Mod. Phys. \textbf{49}, 31 (1977).

\bibitem{Brink}
D. M. Brink and G. R. Satchler, Angular Momentum (Oxford, 1968).



\end{thebibliography}
\end{document}